

\input harvmac

\Title{UCSBTH-91-47}{Correlators of Special States in c=1 Liouville Theory}

\centerline{Miao Li}
\bigskip\centerline{\it Department of Physics}
\centerline{\it University of California}
\centerline{\it Santa Barbara, CA 93106}
\centerline{\it li@voodoo.bitnet}

\vskip .3in

We genaralize the ground ring struture to all other special BRST invariant
operators in the right branch in the $c=1$ Liouville theory.
we also discuss correlation functions of special states on the sphere.

\Date{10/91}%

\newsec{Introduction}

Two dimensional string theory is believed to be a good toy model
for investigating issues such as a complete string field formulation,
gauge symmetries as well as exact blackhole solutions in higher dimensional
strings. Most of all, it is exactly solvable perturbatively \ref\gm{
D.J. Gross and N. Miljkovic, Phys. Lett. 238B (1990) 217; E.Brezin,
V. Kazakov and Al. B. Zamolodchikov, Nucl. Phys. B 338 (1990) 673;
P. Ginsparg and J. Zinn-Justin, Phys. Lett. 240B (1990) 333; G. Parisi,
Phys. Lett. 238B (1990) 209.}. It is remarkable that in addition to
the tachyon field, which should have been the only physical mode
in two dimensions, there are infinitely many discrete physical modes
as remnants of transverse excited fields. These physical states were first
seen in the calculation of the tachyon two point function in the matrix model
\ref\gkn{D.J. Gross, I.R. Klebanov and M.J. Newman, Nucl. Phys. B350
(1991) 621} as poles at special external momenta, and later were confirmed
by Liouville calculation \ref\poly{A.M. Polyakov, Mod. Phys. Lett.
A6 (1991) 635.} and by free field BRST analysis \ref\lz{B. Lian and
G. Zuckerman, Yale preprint YCTP-P18-91; P. Bouwknegt, J. McCarthy
and K. Pilch, CERN preprint CERN-TH.6162/91.}. Much has been done
for correlation functions of tachyons in the matrix model \gkn\
\ref\amp{G. Moore, Yale and Rutgers preprint YCTP-P8-91, RU-91-12;
K. Demeterfi, A. Jevicki and J.P. Rodrigues, Brown preprints
BROWN-HET-795 and BROWN-HET-803 (1991); D.J. Gross and I.R. Klebanov,
Nucl. Phys. B359 (1991) 3.} \ref\ms{G. Moore and N. Seiberg,
Rutgers and Yale preprint RU-91-29, YCTP-P19-91.} and also in
the Liouville theory on the sphere \ref\dk{P. Di Francesco and
D. Kutasov, Phys. Lett. 261B (1991) 385.}. All results are in
remarkable agreement. It remains to be done for correlation functions
of special states and mixed correlation functions of special
states and tachyon states. It is the purpose of this paper to
calculate the former on the sphere in the Liouville theory.

The special states of standard ghost number can be easily seen and constructed
in light of the fact that it has long been known that in c=1 conformal
field theory there are additional primary states at special momenta \ref\kac{
V.G. Kac, in Group Theoretical Methods in Physics, ed. W. Beiglbock
et al. (Springer-Verlag, 1979); G. Segal, Comm. Math. Phys. 80
(1981) 301; R. Dijkgraaf, E. Verlinde and H. Verlinde, Comm. Math.
Phys. 115 (1988) 649.}. Consider a compactified scalar of radius
$\sqrt{2}$. The enlarged current algebra is an $SU(2)$ current algebra of
level 1. The SU(2) currents are $J^{\pm}(z)=\hbox{exp}(\pm i\sqrt{2}X(z))$
and $J^3(z)=i/\sqrt{2}\partial X(z)$. Denote the usual primary field at
momentum $s\sqrt{2}$ by $V_{s,s}=\hbox{exp}(is\sqrt{2}X(z))$, where
$s$ is a positive integer or half integer. Now the additional primary
fields are $V_{s,n}$ obtained by applying the zero mode of $J^{-}$ to
$V_{s,s}$ $s-n$ times. This
construction is valid independent of the radius of $X$. Now the (1,1)
primary field in the Liouville theory is constructed by dressing
$V_{s,n}\bar{V}_{s,n}$ with
$\hbox{exp}(\sqrt{2}(1\mp s)\phi)$, where $\phi$ is the Liouville field.
We shall consider an uncompactified scalar $X$ in which case we have
to have the same momenta in the left and right sectors. Denote, following
Witten \ref\wi
{E. Witten, IAS preprint IASSNS-HEP-91/51.}, these states by $W^{\pm}_{s,n}
=V_{s,n}\hbox{exp}(\sqrt{2}(1\mp s)\phi)$. The corresponding BRST invariant
states are $Y^{\pm}_{s,n}=cW^{\pm}_{s,n}$.

The other special states of non-standard ghost numbers are to be considered
as companions of $Y^{\pm}_{s,n}$, as far as the so-called relative
cohomology is concerned \lz. The companion of $Y^{+}_{s,n}$ has ghost number
zero and we denote it by ${\cal O}_{s-1,n}$. The companion of $Y^-_{s,n}$
has ghost number two and will not concern us in this paper. The $Y^-$'s
do not satisfy Seiberg's condition of a microscopic state \ref\sei
{N. Seiberg, Notes on Quantum Liouville Theory and Quantum Gravity, in
Common Trends in Mathematics and Quantum Field Theory, Procedings of
the 1990 Yukawa International Seminar, ed. T Eguchi et al.}, and are
important in the 2d stringy blackhole \ref\black{E. Witten, IAS preprint
IASSNS-91/12.}. All other BRST
invariant states are irrelevant as far as correlation functions are
concerned \foot{ I am grateful to K. Li for an enlightening discussion on this
issue.}.

Let ${\cal Y}^\pm_{s,n}$ stand for the symmetric left and right combination
$Y^\pm_{s,n}\bar{Y}^\pm_{s,n}$, and ${\cal W}^\pm_{s,n}$ for $W^\pm_{s,n}
\bar{W}^\pm_{s,n}$. First of all, we are interested in calculating correlation
functions on the sphere of the following form
\eqn\corr{\langle {\cal Y}^+_{s_1,n_1}{\cal Y}^+_{s_2,n_2}{\cal Y}
^+_{s_3,n_3}\int {\cal W}^+_{s_4,n_4}
\cdots \int {\cal W}^+_{s_N,n_N}\rangle_\mu .}
Positions of three operators are fixed, so we use ${\cal Y}^+$ instead of
${\cal W}^+$. The above correlator is symmetric in indics $(s_i.n_i)$ by
construction. The subscript $\mu$ indicates the expectation value is defined
with a cosmological term in the action, as in (3.1). One
may further insert a number of operators $\cal {O}$. Since these operators
are of conformal dimension $(0,0)$ and have ghost number zero, the conservation
of ghost number will not be violated and the correlator is independent of
the positions of insertions of $\cal{O}$. One of our main results is that
the correlation in \corr\ is always zero when all $n_i=0$. We conjecture
that all the Liouville bulk correlators in \corr\ are zero. Throughout
this paper all correlators are defined on the sphere.

Witten recently discovered a ground ring structure among operators $\cal{O}$
\wi, and one naturally expects this ring structure be helpful in determining
correlator \corr. This turns out to be partly true. We shall generalize
the ground rign structure to all BRST invariant operators in the
right branch in section 2. What we will show is that the generators of the
ground ring together with a couple of other operators generate all the
special operators in the right branch. We then use this fact to
devise a reshuffling argument in section 4. We will show by this
reshuffling argument that correlator \corr\ when $n_i=0$ can be reduced
to a simple one which depends only on the number of inserted operators
and the sum of $s_i$. And this simple correlator turns out to be zero if
we rescale all operators by a infinitely small factor. We have not been
able to generalize the reshuffling argument to a general case.
Section 3 is devoted to some explicit calculations of three point
correlators of the type in \corr\ and mixed type, namely with
some ${\cal Y}^-$ operators. We use a certain regularization in calculating
correlators. The fact that some correlators are regularized to be zero
is confirmed in section 4 by use of the reshuffling argument.

Some un-rescaled three point correlators of one ${\cal Y}^-$ operator
and two ${\cal Y}^+$ operators calculated in section 3 are relevant
to the perturbed OPE considered in \ref\kp{I.R. Klebanov and A.M.
Polyakov, Princeton preprint, Sept. 1991.}.

We should stress that all correlators (amplitudes) considered in this paper
are Liouville bulk correlators. The discrepancy between our result about
the correlators of the ${\cal Y}^+$'s and the matrix model calculations
in \ref\gkd {D.J. Gross and I.R. Klebanov, Nucl. Phys. B352 (1991) 671; U.H.
Danielsson and D.J. Gross, Princeton preprint PUPT-1258 (1991).} indicates
that the nonvanishing results in the matrix model are to be explained
by the tachyon wall effects.

\newsec{Generalization of The Ground Ring}

Lian and Zuckerman calculated the free field BRST cohomology in the c=1
Liouville theory \lz. In addition to the usual tachyon operators, they
find discrete operators with various ghost numbers. We shall use superscript
$+$ to indicate a discrete operator with a Liouville momentum $p<\sqrt{2}$,
an operator in the right branch; and use superscript $-$ to indicate an
operator with a Liouville momentum $p\ge\sqrt{2}$, an operator in the wrong
branch. As we showed in the introduction, one can construct
operators $Y^+_{s,n}$ and $Y^-_{s,n}$ explicitly. These operators have ghost
number one, the standard ghost number for a BRST invariant operator.

We now briefly recall the main results about special operators in \lz. The
minus operators are simpler. They do not involve oscillators of the Liouville
field. In the relative cohomology which consists of states annihilated
by the zero mode of the anti-ghost $b$, there are operators of ghost number
one and two. The ghost number one operators are just $Y^-_{s,n}$. We will
not be interested in the ghost number two operators. According to \lz,
associated with each operator in the relative cohomology, there is an
operator in the absolute cohomology with a ghost number increased by one.
To be more precise, this operator is not annihilated by $b_0$.
We see that the only ghost number one operators are $Y^-_{s,n}$. As for
the plus opereators, there are operators of ghost number zero and one
in the relative cohomolgy. The ghost number one operators are $Y^+_{s,n}$.
There are no explicit formulas for operators of ghost number zero. However,
Witten discovered \wi\ a ring structure for these operators which we shall
describe presently. Therefore one needs not know the detailed formulas.
Again, associated with these operators are operators of ghost number one
which are not annihilated by $b_0$. The ghost number two counterpart
of $Y^+_{s,n}$ is just $\partial cY^+_{s,n}$. The above is the complete
list of the special operators in the right branch.

Consider operators with ghost number zero. Following Witten \wi, we denote
them by ${\cal O}_{s,n}$. The momenta of ${\cal O}_{s,n}$ are
$\sqrt{2}(n,-s)$. $s$ is an positive half integer or integer, $n=-s,
-s+1,\cdots, s$. Note that instead of using a pure imaginary Liouville
momentum, we use the real Liouville momentum convention. Now the ring
structure of these ${\cal O}$ operators
is obvious. The product ${\cal O}(z){\cal O}'(0)$ is again BRST invariant.
The singular terms in the product must be BRST exact, since there are no
BRST nontrivial operators with negative dimensions. The limit $z\rightarrow
0$ is well defined and the resulting operator must be a third ${\cal O}$
operator, since there are no other operators with ghost number zero. The
ring so defined is a commutative ring and called the ground ring
in \wi. Witten further showed that
${\cal O}_{s,n}$ can be written in a form $x^{s+n}y^{s-n}$. $x$ and $y$
are defined as
\eqn\gen{\eqalign{& x={\cal O}_{{1\over 2},{1\over 2}}=\left(cb+{i\over
\sqrt{2}}(\partial X-i\partial\phi)\right)e^{(iX-\phi)/\sqrt{2}}\cr
&y={\cal O}_{{1\over 2},-{1\over 2}}=\left((cb-{i\over\sqrt{2}}(\partial
X+i\partial\phi)\right)e^{(-iX-\phi)/\sqrt{2}}.}}

A natural question is how to use this ring structure in calculating
correlation functions. To answer this question, one must first generalize
this ring structure to the whole ring of physical operators in the right
branch. To fix notation, we use the following normalization for
$W^+_{s,n}$
\eqn\norm{W^+_{s,n}=(-1)^{2n+1} {1\over 2}(s+n)![(s-n)!]^{1/2}
\left[{1\over 2\pi}\oint J^-\right]^{s-n}e^{i\sqrt{2}sX+\sqrt{2}
(1-s)\phi},}
where the meaning of the contour integrals is the following. The first
contour integral surrounds the position of $W^+_{s,n}$. The next contour
integral surrounds the first one, and so forth. The normalization in
\norm\ is motivated by an explicit calculation of Klebanov and Polyakov
\kp.
Define operators
$$Q^+_{s,n}={1\over 2\pi i}\oint W^+_{s,n}.$$
Klebanov and Polyakov showed that these operators satisfy $w_\infty$ algebra
\eqn\wal{[Q^+_{s,n},Q^+_{s',n'}]=(ns'-n's)Q^+_{s+s'-1,n+n'},}
or equivalently, the following OPE
\eqn\ope{W^+_{s,n}(z)W^+_{s',n'}(0)=\cdots +{1\over z}(ns'-n's)W^+_{s+s'-1,
n+n'}+\cdots .}

Now consider a product $Y^+_{1,0}x^{s+n-1}y^{s-n-1}$, $|n|<s$. This operator
has the same ghost number and momenta as $Y^+_{s,n}$. One may try to identify
these two operators up to a proportionality constant. Unfortunately, as we
have learned, there is another operator with the same ghost number and momenta
in the absolute cohomology. We call this operator $X^+_{s-1,n}$, the
counterpart of operator ${\cal O}_{s-1,n}$ in the absolute cohomology. The
first example is $X^+_{0,0}=\partial c+{1\over\sqrt{2}}c\partial\phi$,
corresponding to the identity operator. In fact, this operator is BRST
exact in some sense. One can show that
$$X^+_{0,0}(z)=[Q,\phi(z)],$$
where $Q$ is the diffeomorphism BRST operator. $X^+_{0,0}$ is considered
as a nontrivial BRST invariant operator in \lz, since there is no state
in the Fock space corresponding to operator $\phi$.

It is easy to see that the term proportional to $\partial c$ in
the product $X^+_{0,0}x^{s+n}y^{s-n}$ is $(s+1)\partial c x^ny^m$.
The latter is defined as a normal ordered product. By a theorem in \lz,
operator $X^+_{0,0}x^{s+n}y^{s-n}$ is a nontrivial operator and is not
annihilated by $b_0$. We define
\eqn\defi{X^+_{s,n}={1\over s+1}X^+_{0,0}x^{s+n}y^{s-n}.}

The product $Y^+_{1,0}x^{s+n-1}y^{s-n-1}$ may contain a term
proportional to $X^+_{s-1,n}$. However, we claim that
\eqn\res{Y^+_{s,n}=(sY^+_{1,0}+nX^+_{0,0})x^{s+n-1}y^{s-n-1},}
where $Y^+_{s,n}=cW^+_{s,n}$ and $W^+_{s,n}$ is defined in \norm.

We proceed to prove \res. Consider the product $Y^+_{1/2,1/2}x^{s+n-1}
y^{s-n}$. It is shown in \wi\ that $Q^+_{1/2,1/2}$ acts on the ground
ring as the differential operator $1/2\partial y$. Using this fact, it is
easy to see that the term proportional to $\partial c$ in the product
under consideration is $(s-n)/2\partial cx^{s+n-1}y^{s-n-1}$. Therefore
we have the following relation
\eqn\star{Y^+_{{1\over 2},{1\over 2}}x^{s+n-1}y^{s-n}=a_{s,n}Y^+_{s,n}+
{s-n\over 2}X^+_{s-1,n}.}
It remains for us to determine the coefficient $a_{s,n}$. To this end,
we first show that
\eqn\diff{Q^+_{s,n}={1\over 2}\left((s+n)x^{s+n-1}y^{s-n}\partial_y
-(s-n)x^{s+n}y^{s-n-1}\partial_x\right),}
when $Q^+_{s,n}$ acts on the ground ring. Indeed, the differential
operators on the left hand side of the above equation form the $w_\infty$
algebra, then $Q^+_{s,n}$ are proportional to them with factors $e^{an}$.
Since $Q^+_{1/2,1/2}$ is just $1/2\partial_y$, so $a=0$.

Calculate $Y^+_{1/2,1/2}x^{s+n-1}y^{s-n+1}$, using \star. We find the product
$a_{s,n}Y^+_{s,n}y$ contains a term $(s+n)/2a_{s,n}X^+_{s-1/2,n+1/2}$ by
use of \diff. The product $(s-n)/2X^+_{s-1,n}y$ is simply $(s-n)(2s+1)/(4s)
X^+_{s-1/2,n+1/2}$, by the definition \defi. Since we expect that $Y^+_{1/2,1/2}
x^{s+n-1}y^{s-n+1}$ must contain $(s-n+1)/2X^+_{s-1/2,n+1/2}$ from \star,
we find $a_{s,n}=1/(2s)$. Finally, we obtain
\eqn\sta{Y^+_{{1\over2},{1\over2}}x^{s+n-1}y^{s-n}={1\over 2s}Y^+_{s,n}
+{s-n\over 2}X^+_{s-1,n}.}
\res\ is a easy consequence of \sta. Rewrite
$$Y^+_{1/2,1/2}x^{s+n-1}y
^{s-n}=Y^+_{1/2,1/2}yx^{s+n-1}y^{s-n-1}={1\over 2}(Y^+_{1,0}+X^+_{0,0})
x^{s+n-1}y^{s-n-1}.$$
Using \sta\ again and the definition \defi\ we find \res.

\defi\ and \res\ tell us how those operators of ghost number one are
generated by $x, y$ together with $Y^+_{1,0}$ and $X^+_{0,0}$. The only
operators of ghost number two in the right branch are $\partial cY^+_{c,n}$.
One might think this operator can be obtained by simply multiplying
$\partial c$ to the right hand side of \res. This is wrong, since $\partial c
$ is not a BRST closed operator hence the product is not well defined.
Nevertheless, we expect from \ope\ that $Y^+_{s,n}Y^+_{s',n'}= (ns'-n's)
\partial cY^+_{s+s'-1,n+n'}$. Using \res\ and the basic products
$Y^+_{1,0}Y^+_{1,0}=X^+_{0,0}X^+_{0,0}=0$ and $X^+_{0,0}Y^+_{1,0}=
-Y^+_{1,0}X^+_{0,0}=\partial cY^+_{1,0}$, we find
\eqn\aap{Y^+_{s,n}Y^+_{s',n'}=(ns'-n's)\partial cY^+_{1,0}x^{s+s'+n+n'-2}
y^{s+s'-n-n'-2}.}
The above equation tells us $\partial cY^+_{s,n}=\left(\partial cY^+_{1,0}
\right)x^{s+n-1}y^{s-n-1}$. Products similar to \aap\ are listed below:
\eqn\abp{\eqalign{&Y^+_{s,n}{\cal O}_{s',n'}={s\over s+s'}Y^+_{s+s',n+n'}
+(ns'-n's)X^+_{s+s'-1,n+n'}\cr
&X^+_{s,n}{\cal O}_{s',n'}={s+s'+1\over s+1}X^+_{s+s',n+n,}\cr
& Y^+_{s,n}X^+_{s',n'}=-{s\over s'+1}\partial cY^+_{s+s',n+n'}\cr
&X^+_{s,n}X^+_{s',n'}=0.}}

We now turn to the ghost number one operators in the wrong branch, namely
$Y^-_{s,n}$. Unfortunately in this case it is not possible to generate all
operators by a single $Y^-$ operator and $x, y$. The reason is simple.
Multiplying $Y^-_{s,n}$ by $x$ or $y$ lowers $s$ to $s-1/2$. While one may
use $Y^-_{s,0}x^ny^m$ to generate another $Y^-$. This product, if not zero,
will be proportional to $Y^-_{s-(n+m)/2,(n-m)/2}$, since there is no other
ghost number one operator with the same momenta in the wrong branch. We shall
show this is indeed the case. First of all, we shall argue that the product
is zero whenever the condition $|(n-m)/2|\le s-(n+m)/2$ is not satisfied.
Conisder first the case when the inequality is saturated and $s-(n+m)/2>0$.
The product is proportional to $Y^-_{|n-m|/2,\pm|n-m|/2}$. It is obvious
that any further multiplication of $x$ or $y$ which makes the
inequality violated will annihilate the operator. Next consider the case
when the inequality is saturated and $s-(n+m)/2=0$. We get an operator
proportional to the cosmological operator $\hbox{exp}(\sqrt{2}\phi)$. It is
easy to see from \gen\ that this operator destroyes $x$ and $y$. In other
words, one can not get an $Y^+$ operator from a product $Y^-_{s,0}x^ny^m$.

To fix the normalization of $Y^-_{s,n}$, again we follow \kp. Define
\eqn\wm{W^-_{s,n}=(-1)^{2n+1}{2\over (2s)!}[(s-n)!]^{-1/2}\left(
{1\over 2\pi i}\oint J^-\right)^{s-n}e^{i\sqrt{2}sX+\sqrt{2}(1+s)\phi}.}
It was shown in \kp\ that
\eqn\aop{W^+_{s,n}(z)W^-_{s',n'}=\cdots -{1\over z}(ns'+n's+n)W^-_{s'-s+1,
n+n'}+\cdots.}

We shall use \aop\ to prove that $Y^-_{s,0}x^ny^m$ is not zero if
$|(n-m)/2|\le s-(n+m)/2$. We divide our proof into three steps.

First consider $Y^-_{s,s}y$ and $Y^-_{s,-s}x$. These opearators must
be proportional to $Y^-_{s-1/2,s-1/2}$ and $Y^-_{s-1/2,-(s-1/2)}$
respectively, assuming they are not zero. Using formula \wm\ and the
definition in \gen, it is fairly easy to show that $Y^-_{s,s}y=-Y^-
_{s-1/2,s-1/2}$ and $Y^-_{s,-s}x=-Y^-_{s-1/2,-(s-1/2)}$. This in
particular implies that
$Y^-_{s,s}y^{2s-1}=(-1)^{2s-1}Y^-_{1/2,1/2}$, which of course is nonvanishing.

Next we prove $Y^-_{s,0}y^m$ is not zero when $m\le s$. It suffices to
prove $Y^-_{s,0}y^s$ not be zero. Consider the commutator
$[Q_{(s+1)/2,(s+1)/2}, Y^-_{s,0}y^s]$. Using the OPE in \aop\ and the
differential operator realization of $Q$ operators when acting on the
ground ring, we obtain
\eqn\key{[Q_{(s+1)/2,(s+1)/2},Y^-_{s,0}y^s]=-{1\over 2}(s+1)^2Y^-_{(s+1)/2,
(s+1)/2}y^s+{s(s+1)\over 2}Y^-_{s,0}x^sy^{s-1}.}
The first term on the r.h.s. is not zero, as we learned before. Suppose
$Y^-_{s,0}y^s$ vanishes. Then the second term on the r.h.s. must
be nonzero, in order to cancel the first term. Therefore,
$Y^-_{s,0}x^sy^{s-1}\ne 0$. This implies $Y^-_{s,0}x^s\ne 0$. However,
we know the symmetry between $x$ and $y$ under reflection of the matter
field $X\rightarrow -X$, $Y^-_{s,0}y^s$ must be non-vanishing too. So the
only consistent solution is for both of them to be non-zero.

It remains to show that $Y^-_{s,0}x^ny^m$ is not zero. Without loss of
generality, we can assume $m>n$ (The case $n>m$ can be similarly considered.
The $n=m$
case is a consequence of $n\ne m$ cases). Now the only condition is
$m-n\le s$. Consider the commutator $[Q_{n+1,0}, Y^-_{s,0}y^{m-n}]$. It must be
nonzero, as indicated by \aop. Note that $Q_{n+1,0}$ commutes with $Y^-_{
s,0}$. We then find
\eqn\ke{[Q_{n+1,0}, Y^-_{s,0}y^{m-n}]={m-n\over 2}(n+1)Y^-_{s,0}x^ny^m.}
The immediate consequence of the above equation is that $Y^-_{s,0}x^ny^m$
must be non-vanishing. After a few calculations, we find
\eqn\for{Y^-_{s,n}=(-1)^{2n}Y^-_{S,0}x^{S+n-s}y^{S-n-s},}
where $S$ is an arbitrary integer such that $S\pm n-s\ge 0$.

\newsec{Three Point Correlators of Special Operators}

We shall calculate various three point functions of special states in this
section. The action under consiedration is
\eqn\act{S={1\over 2\pi}\int\left(\partial X\overline{\partial}X+
\partial\phi\overline{\partial}\phi-{1\over\sqrt{2}}R\phi+2\mu
\Gamma(\epsilon)e^{\sqrt{2}\phi}\right),}
where we have suppressed the action of the ghosts. We have put a short
distance cut-off $\Gamma(\epsilon)$, $\epsilon
\rightarrow 0^+$ explicitly into the cosmological term. The origin of this
cut-off is of course the short distance divergence. In calculating an
amplitude of tachyons, one needs to perform a multi-complex integral. The
integrand usually displays poles when two complex points approach to
each other. There are two equivalent regularization prescriptions. One
prescription is
to simply introduce a cut-off in the integral such that any two of complex
positions are saparated by a mimimal distance. Another is to shift
the exponent in each factor $|z_i-z_j|^{\alpha_{ij}}$ by a small amount
proportional to $\epsilon$. It turns out that $\Gamma(\epsilon)$ will
appear in the place of $|\hbox{log}\lambda|$, $\lambda$ is the short distance
cut-off. The reason for us to put a factor $\Gamma(\epsilon)$ in the
cosmological
term is that the cosmological term decouples from the tachyon amplitude
thereby needs an infinity rescaling factor. This is true for those special
tachyon vertices in the wrong branch \dk, namely when the Liouville momentum
$p\ge\sqrt{2}$. For those special tachyon vertices in the right branch,
an infinitely small factor $[\Gamma(\epsilon)]^{-1}$ is needed in rescaling
vertices, however. One expects this phenomenon persist for special operators
we are considering.

We are going to use the same trick as in
\ref\gl{A. Gupta, S.P. Trivedi and M.B. Wise, Nucl. Phys. B340
(1990) 475; M. Goulian and M. Li, Phys. Rev. Lett. 66 (1991) 2051.}
to calculate three point correlators. We first perform integration over
the Liouville zero mode, effectively bringing down a power of the cosmological
term. The power is always a positive integer for three ${\cal Y}^+$ operators.
So this ad hoc trick is equivalent to expanding $e^{-S}$ in terms of the
cosmological term. We expect a scaling behavior $(s!)^{-1}(-\mu)^s|\hbox{log}
\mu|$ in this case. The term $|\hbox{log}\mu|$ is just the Liouville
volume, resulting from the integration over the Liouville zero mode.
The power of the cosmological term we bring down may become zero or
a negative integer in other cases. Then we need an analytic continuation
technique used in the second paper in \gl.

\vskip8pt
\noindent {\bf 3.1 Three Point Correlators of ${\cal Y}^+$}
\vskip4pt

We are going to calculate
$$\langle \prod_{i=1}^3{\cal Y}^+_{s_i,n_i}\rangle_\mu,$$
where the subscript $\mu$ indicates the correlator is defined with action
\act. We integrate over the Liouville zero mode to obtain
\eqn\bare{\eqalign{&{1\over s!}(-\mu/\pi)^s|\hbox{log}\mu|\Gamma^s(\epsilon)
\langle\prod_{i=1}^3{\cal Y}^+_{s_i,n_i}\left(\int e^{\sqrt{2}\phi}
\right)^s\rangle\cr
&s=-1+\sum_is_i.}}
The expectation value in \bare\ is to be calculated without the cosmological
term in the action. The integration of the Liouville zero mode is represented
by $|\hbox{log}\mu|$. We have used ${\cal Y}^+_{s_i,n_i}$ instead of ${\cal
W}^+_{s_i,n_i}$ to indicate that positions of these operators are fixed.
To calculate the expectation value in \bare, we may first calculate the
correlation function without integrations of the cosmological term, also let
positions of ${\cal W}^+_{s_i,n_i}$ not be fixed. Let $z_i$ be positions
of ${\cal W}^+$ and $w_i$ positions of $e^{\sqrt{2}\phi}$. Now the part
from the matter sector $X$ is just
$$\eqalign{&C_{s_1,n_1,s_2,n_2}^{s_3,n_3}\prod_{i<j}^3|z_i-z_j|^{-2
\Delta_{ij}}\cr
&\Delta_{12}=s^2_1+s^2_2-s^2_3,}$$
other $\Delta_{ij}$ are obtained from $\Delta_{12}$ by permutations. The
structure constant $C$ is proportional to the Clebsch-Gordon coefficient.
It is not zero only when $\sum_i n_i=0$ and $s_3=|s_1-s_2|,\cdots, s_1+s_2$.
Next we calculate the contribution from the Liouville sector. The result
is
$$\prod_{i<j}^3|z_i-z_j|^{-4(1-s_i)(1-s_j)}\prod_{i,j}|w_i-z_j|^{-4(1-s_j)}
\prod_{i<j}^s|w_i-w_j|^{-4}.$$

Our strategy is to fix $z_i$ and integrate over $w_i$. We immediately find
that the multi-complex integral is divergent. To regularize the integral,
we should shift the expononets in $|w_i-w_j|^{-4}$ to $-4+4\epsilon$. To
have a $SL(2,C)$ invariant integrant, other exponents must be shifted
accordingly. For example, the exponent in $|w_i-z_j|^{-4(1-s_j)}$ is shifted
to $-4(1-s_j)-4\epsilon (s_j-2/3)$. After doing shifting and fixing $z_i$,
the integral we need to do is
\eqn\intg{\int \prod_i^sd^2w_i|w_i|^{-4(1-s_1)-4\epsilon(s_1-2/3)}
|1-w_i|^{-4(1-s_2)-4\epsilon(s_2-2/3)}\prod_{i<j}^s|w_i-w_j|^{-4+4\epsilon}.}
This integral is not symmetric in all $s_i$ superficially. As we soon see,
the final result is symmetric. The integral is performed in \ref\df{Vl.S.
Dotsenko and V.A. Fateev, Nucl. Phys. B251 (1985) 691.}. We simply write
down the final answer
\eqn\answ{\eqalign{&(-1)^{1+s}{\pi^s\over (s!)^2}\left({\Gamma(1/3)\over
\Gamma(2/3)}\right)^3\prod_{j=1}^s\Gamma^{-8}(j)\cr
&\prod_{i=1}^3\left(
{\Gamma(2s_i-1/3)\over\Gamma(s+1-2s_i+1/3)}\prod_{j=1}^{2s_i-1}\Gamma^2(s+1
-j)\Gamma^2(j)\right)\Gamma^{2-s}(\epsilon).}}
Note that there is a factor $\Gamma^{-s}(\epsilon)$ appearing in the above
formula and is to be cancelled by the factor $\Gamma^s(\epsilon)$ in \bare.
Note also that we have assumed that each $s_i\ge 1$, so $s\ge 2$. The above
result tells us that the multi-complex integral goes to zero in general.

Collecting \bare\ and \answ\ together, we find
\eqn\plus{\eqalign{&\langle\prod_{i=1}^3{\cal Y}^+_{s_i,n_i}\rangle_\mu
=-\mu^s|\hbox{log}\mu|F(s_i,n_i)\Gamma^2(\epsilon)\cr
&F(s_i,n_i)=C_{s_1,n_1,s_2,n_2}^{s_3,n_3}\left({\Gamma(1/3)\over\Gamma
(2/3)s!}\right)^3\prod_{j=1}^s\Gamma^{-8}(j)\cr
&\qquad \prod_{i=1}^3\left({\Gamma(2s_i-1/3)\over\Gamma(s+1-2s_i+1/3)}
\prod_{j=1}^{2s_i-1}\Gamma^2(s+1-j)\Gamma^2(j)\right).}}
The final result is divergent, as we expected. To have a finite result,
one should rescale operators ${\cal W}^+$. If one rescale each operator
by a factor $\Gamma^{-1}(\epsilon)$, then one has a vanishing result
from \plus. We can not attribute the factor $\Gamma^{-1}(\epsilon)$
in the rescaled correlator to the partition function. The reason is the
following. The correlators of tachyon vertices of generic momenta are
finite, no matter how one rescale the partition function. To agree with
the matrix model calculation, one therefore requires a finite partition
function.

\vskip8pt
\noindent {\bf 3.2 Three Point Correlators of Mixed Type}
\vskip4pt
Next we calculate three point correlators of mixed type, i.e. with
presence of both ${\cal Y}^+$ and ${\cal Y}^-$ operators. First, consider
correlators of two ${\cal Y}^+$ with one ${\cal Y}^-$. We shall calculate
$$\langle {\cal Y}^+_{s_1,n_1}{\cal Y}^+_{s_2,n_2}{\cal Y}^-_{s_3,n_3}\rangle
_\mu.$$

Again we perform the integration over the Liouville zero mode first, we will
get a formula similar to \bare. Now $s=s_1+s_2-s_3-1$, which can be positive,
zero and negative. We would have exactly same formula as in \bare\ if $s$ is
positive. The structure constant $C$ must be replaced by
another $\tilde{C}$, since we use different normalization for the
matter part of ${\cal Y}^-$ (see section 2). We repeat steps in the previous
calculations and
finally reach \intg. The integral is not given by \answ\ since this time
$s_3$ appears in $s$ with a different sign. It is not hard to use a formula
in \df\ to obtain the regularized integral
\eqn\ana{\eqalign{&(-\pi)^s\prod_{i=1}^s\Gamma^2(2s_1-i)\Gamma^2(2s_2-i)
\Gamma^{-2}(2s_3+i+1)\Gamma^{-2}(i+1)\cr
&{\Gamma(2s_1-1/3)\Gamma(2s_2-1/3)\Gamma(2s_3+4/3)\over \Gamma(2s_1-s-1/3)
\Gamma(2s_2-s-1/3)\Gamma(2s_3+s+4/3)}\Gamma^{-s}(\epsilon).}}
The arguments in gamma functions in the product are all positive, since for
example $2s_1-s=s_1+s_3-s_2+1$ is always positive by the fusion rules in the
matter sector. It is notable that the integral goes to zero for positive
$s$ for which \ana\ is valid. Note that the above formula is also good for
$s=0$, provided we forget about the product $\prod_{i=1}^s$. It is just 1 as
expected. We write down the correlator
\eqn\mix{\eqalign{&\langle{\cal Y}^+_{s_1,n_1}{\cal Y}^+_{s_2,n_2}{\cal
Y}^-_{s_3,n_3}\rangle_\mu={\mu^s\over s!}|\hbox{log}\mu|G(s_i,n_i)\cr
&G(s_i,n_i)=\tilde{C}_{s_1,n_1,s_2,n_2}^{s_3,n_3}{\Gamma(2s_1-1/3)
\Gamma(2s_2-1/3)\Gamma(2s_3+4/3)\over
\Gamma(2s_1-s-1/3)\Gamma(2s_2-s-1/3)\Gamma(2s_3+s+4/3)}\cr
&\prod_{i=1}^s\Gamma^2(2s_1-i)\Gamma^2(2s_2-i)\Gamma^{-2}(2s_3+i+1)
\Gamma^{-2}(i+1).}}
The regulator $\Gamma(\epsilon)$ simply disappears in the correlator.

There are two implications of formula \mix. First, as one should rescale
${\cal Y}^+$ operators by $\Gamma^{-1}(\epsilon)$, one should also rescale
${\cal Y}^-$ operators by $\Gamma(\epsilon)$. We already know that this is
necessary for the cosmological term. If we do so, the rescaled correlator
will be zero again, as \mix\ tells us. However, if one does not rescale
any operator at all, we will get perturbed OPE of two ${\cal W}^+$ operators
from \mix. The $s=0$ case corresponds to the leading term in the OPE and
is given in \kp. It is just the unperturbed OPE. Thus it would be interesting
to use \mix\ to check the conjecture made in \kp\ about the full OPE and to
construct a space-time action for special states with a nonzero cosmological
constant $\mu$.

The negative $s$ is unique, $s=-1$. This is because the maximal value of
$s_3$ is $s_1+s_2$. To define the ``integral'', we need use a kind of
analytic continuation. Indeed every term in \ana\ is perfectly defined
except for those in the product. The product is of form $\prod^{-1}_{i=1}
f(i)$. If we are bold enough to use Dotsenko's suggestion \ref\dot{V.S.
Dotsenko, preprint PAR-LPTHE 91-18.} to define a product $\prod^{-n-1}
_{i=1}f(i)$ as $\prod^{n}_{i=0}f^{-1}(-i)$, we can rewrite the
un-defined product
in \ana. There is a subtlety to notice, however. If we start with the
formula in \df\ for the integral and use Dotsenko's prescription, the result
is different from the one we can read from \ana. The correct answer is
\eqn\fin{(-\pi)^{-1}{\Gamma^2(2s_3+1)(2s_3+1/3)\over \Gamma^2(2s_1)\Gamma^2
(2s_2)(2s_1-1/3)(2s_2-1/3)}\Gamma^2(\epsilon).}
The result is divergent as $\Gamma^2(\epsilon)$. Had we start with \ana,
we would have obtained a divergent factor $\Gamma(\epsilon)$ and the
same numeric coefficient as in \fin. Using \fin\ we obtain the three
point correlator
\eqn\thl{\eqalign{&\langle {\cal Y}^+_{s_1,n_1}{\cal Y}^+_{s_2,n_2}
{\cal Y}^-_{s_1+s_2,n_3}\rangle_\mu=\tilde{C}^{s_1+s_2,n_3}_{s_1,n_1,s_2,n_2}
\cr
&\mu^{-1}{\Gamma^2(2s_3+1)
(2s_3+1/3)\over \Gamma^2(2s_1)\Gamma^2(2s_2)(2s_1-1/3)(2s_2-1/3)}
\Gamma(\epsilon).}}
The integration of the Liouville zero mode gives $\Gamma(1)=1$ instead of
the Liouville volume $|\hbox{log}\mu|$. Unlike \mix, the correlator in \thl\
is divergent and the rescaled correlator is finite.

\newsec{A Reshuffling Argument}

We have learned from \answ\ that the ``bare'' expectation value in \bare\
is zero if $s>2$. This is the generic case since we consider $s_i\ge 1$
only. It would be nice if we can confirm this result by using the product
representation of ${\cal Y}^+$ operators in section 2. We shall use a
reshuffling argument which unfortunately applies to three point correlators
only thus far. Let us consider a four point correlator of the form
$$\langle{\cal O}_{s_0,n_0}\overline{{\cal O}}_{s_0,n_0}\prod_{i=1}^3
{\cal Y}^+_{s_i,n_i}\rangle_\mu,$$
where we have inserted another operator ${\cal O}\overline{{\cal O}}$.
All positions of operators in the correlator are fixed, since all operators are
of conformal dimension $(0,0)$ and the total ghost number in each sector
is three.

Again we integrate the Liouville zero mode first and obtain
\eqn\four{{1\over s!}(-\mu/\pi)^s|\hbox{log}\mu|\Gamma^s(\epsilon)\langle
{\cal O}_{s_0,n_0}\overline{{\cal O}}_{s_0,n_0}\prod_{i=1}^3{\cal Y}^+_{
s_i,n_i}\left(\int e^{\sqrt{2}\phi}\right)^s\rangle.}
We have assumed $s=-1\sum_{i=0}^3s_i$ be positive. The expectation value
in \four\ is again
the ``bare'' expectation value, without the cosmological term in the action.
Now our reshulling argument goes as follows. Since ${\cal O}\overline
{{\cal O}}$ is BRST invariant, the expectation value in \four\ should
not depend on the
position of its insertion. We move the position and let it hit any of
three ${\cal Y}^+$ operators. The resulting operator of the product
of this ${\cal Y}^+$ operator with ${\cal O}\overline{{\cal O}}$ is not
another ${\cal Y}^+$ in general, as formula \res\ in section 2 tells us.
The resulting operator is rather a sum of one ${\cal Y}^+$ operator and
one operator ${\cal X}^+=X^+\overline{X}^+$. In any case, this operation
tells us one can move $x$ and $y$ (together with $\bar{x}$ and $\bar{y}$
in the anti-holomorphic sector) from one ${\cal Y}^+$ to another. This is
our resuffling argument. One may worry about contact terms caused by
integrations of the cosmological term in \four. As we learned in section
2, the cosmological term kills other operators when one considers a product
with its presence. Therefore the contact terms are zero.

The main difficulties in generalizing the resuffling argument to a general
correlator are obvious. Some positions of ${\cal Y}^+$ operators must
be integrated. One can not reshuffle $x$ and $y$ factors among all ${\cal
Y}^+$ operators, since some reshuffling will result in a ${\cal X}^+$
operator and integration of this operator over the surface is not well
defined \ref\pn{J. Polchinski, Nucl. Phys. B307 (1988) 61; P. Nelson,
Phys. Rev. Lett. 62 (1989) 993.}. Another issue is that one should worry
about contact terms. These difficulties disappear if we consider
correlators of operators of zero momenta in the $X$ direction exclusively.
We shall discuss these correlators in the end of this section.

Come back to the three point correlator, we would like to show that it is
zero by using the reshuffling argument. From \plus\ we learn that the
dependence of the bare correlator on $n_i$ is through the structure
constant $C$. One can always choose $n_1=0$ and a certain $n_2=-n_3=n$
such that the structure constant is not zero (there is always an integer
$s_i$, let $s_1$ be an integer). To prove the vanishing of
the bare correlator, we assume $n_1=0$. By use of the reshuffling argument
and \res\ in section 2, we get
\eqn\any{\eqalign{&\langle Y^+_{s_1,0}Y^+_{s_2,n}Y^+_{s_3,-n}\left(\int
e^{\sqrt{2}\phi}\right)^s\rangle =\cr
&s_1\langle (Y^+_{1,0}x^{s-2}y^{s-2})(s_2Y^+_{1,0}+n
X^+_{0,0})(s_3Y^+_{1,0}-nX^+_{0,0})\left(\int e^{\sqrt{2}\phi}\right)^s
\rangle,}}
where we omitted the anti-holomorphic sector. Now $Y^+_{1,0}x^{s-2}y^{s-2}$
is proportional to $Y^+_{s-1,0}$. The matter part in $Y^+_{s-1,0}$ is
proportional to
$$\left(\oint J^-\right)^{s-1}\hbox{exp}(i\sqrt{2}(s-1)X).$$
It is easy to see that for the matter part on the right hand side of \any,
one needs only to evaluate correlators
\eqn\don{\eqalign{&\langle[\left(\oint J^-\right)^{s-1}e^{i\sqrt{2}(s-1)X}]
\partial X\partial X\rangle\cr
&\langle [\left(\oint J^-\right)^{s-1}e^{i\sqrt{2}(s-1)X}]\partial X\rangle.}}
In both correlators, there are a number of insertions of contour integral. We
can now deform these contour integrals to act on $\partial X$. Note that
$$\oint J^-\partial X\propto \hbox{exp}(-i\sqrt{2}X)$$
and $\left(\oint J^-\right)^2\partial X=0$, we conclude that these correlators
in \don\ are zero if $s>3$. When $s=3$, the first correlator is not zero,
one needs to go back to calculations in the previous section. When $s=2$,
both correlators are not zero. Indeed in \answ\ the multi-complex
integral is not zero for $s=2$.

A remark is in order. One may wonder why we do not just calculate the OPE
in \any\ to show directly that the correlator is zero, uisng the fact that
$Y^+_{1,0}Y^+_{1,0}=X^+_{0,0}X^+_{0,0}=0$. The answer is the following.
These products are zero up to BRST commutators, for example $Y^+_{1,0}
Y^+_{1,0}=
[Q, -1/2\partial c]$. Since $\partial c$ is not annihilated by $b_0$, the
usual argument for decoupling of a BRST commutator does not go through.

We use the same argument to show that the bare correlator of two
${\cal Y}^+$ operators and one ${\cal Y}^-$ operator is zero when
$s>0$, in agreement with \ana. Choose the representation $Y^-_{s_3,n_3}
=(-1)^{2n_3}Y^-_{S,0}x^{S+n_3,-s_3}y^{S-n_3-s_3}$ as in section 2.
Applying the reshuffling argument, we find
\eqn\pol{\eqalign{&\langle Y^-_{s_3,n_3}Y^+_{s_1,n_2}Y^+_{s_2,n_2}\left(
\int e^{\sqrt{2}\phi}\right)^s=\cr
&(-1)^{2n_3}\langle (Y^-_{S,0}x^{S+s-1}y^{S+s-1})(s_1Y^+_{1,0}+n_1X^+_{0,0})
(s_2Y^+_{1,0}+n_2X^+_{0,0})\left(e^{\sqrt{2}\phi}\right)^s\rangle,}}
where we suppressed the anti-holomorphic sector again. As we showed in section
2, the product $Y^-_{S,0}x^{S+s-1}y^{S+s-1}$ is simply zero when $s>1$. When
$s=1$, the product is just the cosmological term. One should calculate the
multi-complex integral again. It is interesting to note that the argument
does not force the bare correlator be zero when $s=0,-1$, as we know it is
not zero for these values. Applying the reshuffling argument, we will
find that bare correlators of two ${\cal Y}^-$ operators and one ${\cal Y}^+$
operator are zero.

Finally we consider correlators of ${\cal Y}^+$ operators with zero $X$
momentum. We do not have to worry about ${\cal X}^+$ operators since they
do not appear here. The only things worrying us are contact terms. It is easy
to see that the contact term of two such ${\cal Y}^+$ operators is zero.
With the resuffling argument, the following correlator
\eqn\zero{\eqalign{&\langle \prod_{i=1}^3{\cal Y}^+_{s_i,0}\prod_{i=4}^N
\int{\cal W}^+_{s_i,0}\left(\int e^{\sqrt{2}\phi}\right)^s\rangle\cr
&s=2+\sum_{i=1}^n(s_i-1)}}
is reduced to
\eqn\red{{1\over s-1}\prod_{i=1}^Ns_i\langle{\cal Y}^+_{s-1,0}({\cal Y}^+_{1,0}
)^2\left(\int{\cal W}^+_{1,0}\right)^{N-3}\left(\int e^{\sqrt{2}\phi}
\right)^s\rangle.}
Now the matter part of $Y^+_{s-1,0}$ is proportional to $(\oint J^-)^{s-1}
\hbox{exp}(i\sqrt{2}(s-1)X)$. We can apply the contour deformation to \red\
again. An immediate consequence is that the correlator is zero when
$s-1>N-1$. When $s-1=N-1$, the correlator can be explicitly calculated.
We have to evaluate a regularized multi-complex integral again. We shall
not do it here. Suffices it to say that the regularized integral is
proportional to $\Gamma^{-1}(\epsilon)$, which is just zero. $N=3$ is a
special case, and from \answ\ we see that the regularized integral indeed
goes to zero. It can be proven that even when $s-1<N-1$, the regularized
multi-complex integral is zero.

To summarize, we have shown that for all three point correlators and
correlators of ${\cal Y}^+$ with zero $X$ momentum the bare correlators
are zero whenever the scaling exponent $s>0$. We would like to conjecture
that this is always the case and futhermore the rescaled correlators
are zero whenever $s>0$.

\newsec{Conclusion}

We have shown that the generators $x$ and $y$ of the ground ring can be
used to generate other BRST invariant operators. This fact
makes possible a reshuffling argument by which some calculations of three
point correlators in section 3 are confirmed by an independent method.
We also show that bare correlators of operators with zero $X$ momentum
vanish. This leads us to conjecture that indeed all Liouville bulk
correlators of operators in the right branch are zero. If this is true,
then the nontrivial results obtained in \gkd\ must be explained in other way,
maybe by tachyon wall effects. It would be interesting to make further use
of the ground ring structure.

\noindent {\bf Acknowledgements}

I have benefitted from many invaluable discussions with K. Li. I wish
to thank B. Blok, S. Giddings, N. Ishibashi, A. Steif and A. Strominger
for comments on the manuscript. This work was supported by DOE grant
DOE-76ER70023.

\vfill\eject
\listrefs

\bye